\documentclass[pra,aps,preprint]{revtex4-1}
\usepackage{amsmath}
\usepackage{amssymb}
\usepackage{graphicx}

\newcommand{\mP}{{\mathcal P}}
\newcommand{\mT}{{\mathcal T}}

\newcommand{\beqa}{\begin{eqnarray}}
\newcommand{\eeqa}{\end{eqnarray}}

\begin{document}
\title{The robust $\mP\mT$-symmetric chain and properties of its Hermitian counterpart} 
\author{Yogesh N. Joglekar}
\email{yojoglek@iupui.edu}
\affiliation{Department of Physics, 
Indiana University Purdue University Indianapolis (IUPUI), 
Indianapolis, Indiana 46202, USA}
\author{Avadh Saxena}
\affiliation{Theoretical Division, Los Alamos National Laboratory, Los Alamos, New Mexico 87545, USA}
\date{\today}
\begin{abstract}
We study the properties of a parity- and time-reversal- ($\mP\mT$) symmetric tight-binding chain of size $N$ with position-dependent hopping amplitude. In contrast to the fragile $\mP\mT$-symmetric phase of a chain with constant hopping and imaginary impurity potentials, we show that, under very general conditions,  our model is {\it always} in the $\mP\mT$-symmetric phase. We numerically obtain the energy spectrum and the density of states of such a chain, and show that they are widely tunable. By studying the size-dependence of  inverse participation ratios, we show that although the chain is not translationally invariant, most of its eigenstates are extended. Our results indicate that tight-binding models with non-Hermitian, $\mP\mT$-symmetric hopping have a robust $\mP\mT$-symmetric phase  and rich dynamics which may be explored in coupled waveguides. 
\end{abstract}
\maketitle

\noindent{\it Introduction:} Since the seminal paper by Bender {\it et al.}~\cite{bender} a decade ago, it has become clear that non-Hermitian Hamiltonians with parity and time-reversal ($\mP\mT$)-symmetry can have purely real spectra~\cite{bender2,bender3} and, with an appropriately redefined inner-product, they lead to orthogonal eigenvectors~\cite{bender2}, unitary scattering~\cite{znojilsc} and, therefore, a consistent quantum theory. The theoretical work on continuum, $\mP\mT$-symmetric, non-Hermitian Hamiltonians~\cite{mostafa} since then has been accompanied, most recently, by experiments in optics where spontaneous $\mP\mT$-symmetry breaking in a classical system  has been observed in waveguides with $\mP\mT$-symmetric complex refractive index~\cite{theory,exp} and by theoretical studies of distributed-feedback optical structures that can be mapped onto a relativistic, $\mP\mT$-symmetric Hamiltonian~\cite{longhi}.

Idealized lattice models have been popular in physics due to their analytical and numerical tractability, the absence of divergences~\cite{kogut}, the availability of exact solutions~\cite{onsager}, and the ability to capture counter-intuitive physical phenomena~\cite{winkler}. As with the standard quantum theory, these models have been based on Hermitian Hamiltonians. In recent years, tight-binding models with a Hermitian hopping and $\mP\mT$-symmetric, complex, on-site potentials~\cite{bendix,song,mark}, non-Hermitian transitions~\cite{zno}, and $\mP\mT$-symmetric spin-chains~\cite{pts} have been extensively explored. For a tight-binding chain with $\mP\mT$-symmetric impurity potentials, a salient result is that its $\mP\mT$-symmetric phase - the range of model parameters that lead to a real spectrum - is extremely fragile~\cite{bendix,mark}. This fragile nature of the $\mP\mT$-symmetric phase precludes effects such as the Anderson localization~\cite{anderson}, impurity-bound states~\cite{sasha}, and the Luttinger-liquid behavior \cite{luttinger} in such a chain with a non-Hermitian Hamiltonian. 

In this paper, we explore the properties of a tight-binding chain of size $N$ with $\mP\mT$-symmetric, non-Hermitian, position-dependent hopping amplitudes. Our main results are as follows: (i) We show that the system is {\it always in the $\mP\mT$-symmetric phase} under very general criteria that we derive. (ii) The energy spectrum and the resulting density of states in such a chain are widely tunable and symmetric around zero. (iii) Although the chain is not translationally invariant, (a majority of) its eigenfunctions are delocalized. Our results show that a {\it robust} $\mP\mT$-symmetric chain has non-Hermitian hopping amplitudes and Hermitian potentials, and that its Hamiltonian is similar to that of a chain with position-dependent, parity-symmetric hopping~\cite{gautam}.


\noindent{\it Tight-binding Model:} We start with a Hamiltonian for an $N$-site tight-binding chain,
\begin{equation}
\label{eq:hpt}
H_{PT}=-\sum_{i=1}^{N-1} \left( t_i c^{\dagger}_{i+1} c_i + t^{*}_{N-i} c^{\dagger}_{i} c_{i+1}\right)
\end{equation}
where $c^{\dagger}_n (c_n)$ is the creation (annihilation) operator at site $n$, $t_i$ are the position-dependent hopping amplitudes, and the asterisk denotes complex conjugation. The parity operator on the chain is given by $\langle m| \mP| n\rangle=\delta_{m,N+1-n}=\delta_{m,\bar{n}}$ where $|m\rangle$ represents a single-particle state localized at site $m$, and $\bar{m}=N+1-m$ is the reflection-counterpart of site $m$; it follows that $H_{PT}$, although not Hermitian, is $\mP\mT$-symmetric. We consider only the single-particle sector and, since periodic boundary conditions are incompatible with the $\mP\mT$-symmetry, use open boundary conditions. Numerical results indicate that the spectrum of $H_{PT}$ is purely real when the hopping elements have the same sign; in the following paragraph, we analytically derive the criteria that guarantee this robustness. 

Let us consider a similarity transformation~\cite{mostafa} of the non-Hermitian Hamiltonian, $H_{PT}\rightarrow H=M^{-1}H_{PT} M$ where $M={\rm diag}(m_1,\ldots,m_N)$ is a diagonal matrix. It is straightforward to show that the transformed matrix $H$ is Hermitian, $H=H^{\dagger}$, if and only if 
\begin{equation}
\label{eq:constraint}
\frac{m^{*}_{k+1}m_{k+1}}{m^{*}_km_k}=\left(\frac{t_{N-k}}{t_k}\right)>0. 
\end{equation}
We note that this constraint only applies to a diagonal $M$. Thus, the $\mP\mT$-symmetric Hamiltonian $H_{PT}$ is {\it similar} to a Hermitian Hamiltonian $H$ if and only if the phases of the hopping amplitudes $(t_k, t_{N-k})$ are the same for all $k=\{1,\ldots,N-1\}$. When the hopping elements are real, it implies that $t_k$ and $t_{N-k}$ must have the same sign; when they are complex, $t_m=|t_m|\exp(i\theta_m)$, it implies that $\theta_k=\theta_{N-k}$. {\it The eigenvalue spectrum of the non-Hermitian Hamiltonian $H_{PT}$ is purely real, as long as these general requirements are satisfied.}  Since $H=M^{-1} H_{PT}M=H^{\dagger}$, it follows that the eigenvalues $E_n$ of $H$ and $H_{PT}$ are the same, and that the orthogonal eigenvectors of $H$, $H|v_n\rangle=E_n|v_n\rangle$, and the (non-orthogonal) eigenvectors of the $\mP\mT$-symmetric Hamiltonian, $H_{PT} |u_n\rangle= E_n |u_n\rangle$ are related by $|u_n\rangle=M|v_n\rangle$. This relation provides the requisite inner-product under which the eigenvectors  $|u_n\rangle$ of $H_{PT}$ are orthonormalized. We note this transformation corresponds to the positive-definite, self-adjoint, invertible metric $\eta^{-1}= M M^\dagger$~\cite{pts}. 

The hopping amplitudes for atomic orbitals can be, in general, complex~\cite{slater}. However, for optical lattices, coupled waveguides, or superlattices, the hopping amplitude, determined by the overlap of adjacent on-site (Gaussian or exponential) ground-state wavefunctions, is positive, just as it is for $s$-wave atomic orbitals~\cite{slater}. A truly non-Hermitian Hamiltonian $H_{PT}$ may be realized in systems with asymmetrical hopping due to an in-plane field~\cite{hatano} or a voltage bias~\cite{beloch}. Its Hermitian counterpart $H$, with position-dependent, parity symmetric hopping, may be realized in evanescently coupled waveguides where the wavepacket evolution and two-particle quantum correlations are exquisitely sensitive to the hopping~\cite{gautam}. 

Note that Eq.(\ref{eq:constraint}), although dependent upon the underlying Hamiltonian $H_{PT}$, does not uniquely determine the transformation matrix $M$ or the Hermitian matrix $H$. For simplicity, we choose $M$ to be real and $m_1=1$ which implies, via $m_{k+1}=m_k\sqrt{t_{N-k}/t_k}$, that $m_N=1$. The resulting real matrix $M$ commutes individually with the parity- and time-reversal operators. Since numerical diagonalization of a Hermitian matrix $H$ is faster and more accurate than its non-Hermitian counterpart $H_{PT}$, in numerical calculations we use its Hermitian counterpart $H$ with entries
\begin{equation}
\label{eq:hh}
H_{mn}=-|t_m t_{N-m}|^{1/2}\left(\delta_{m,n-1}e^{i\theta_m} +\delta_{m-1,n}e^{-i\theta_{N-n}}\right),
\end{equation}
where we recall that $\theta_{N-n}=\theta_n$. In the following sections, we discuss basic properties of such a chain, with focus on the energy spectrum and nature of wavefunctions of $H$ when the hopping is not uniform. 

\begin{figure}[h!]
\begin{center}
\includegraphics[angle=0,width=17cm]{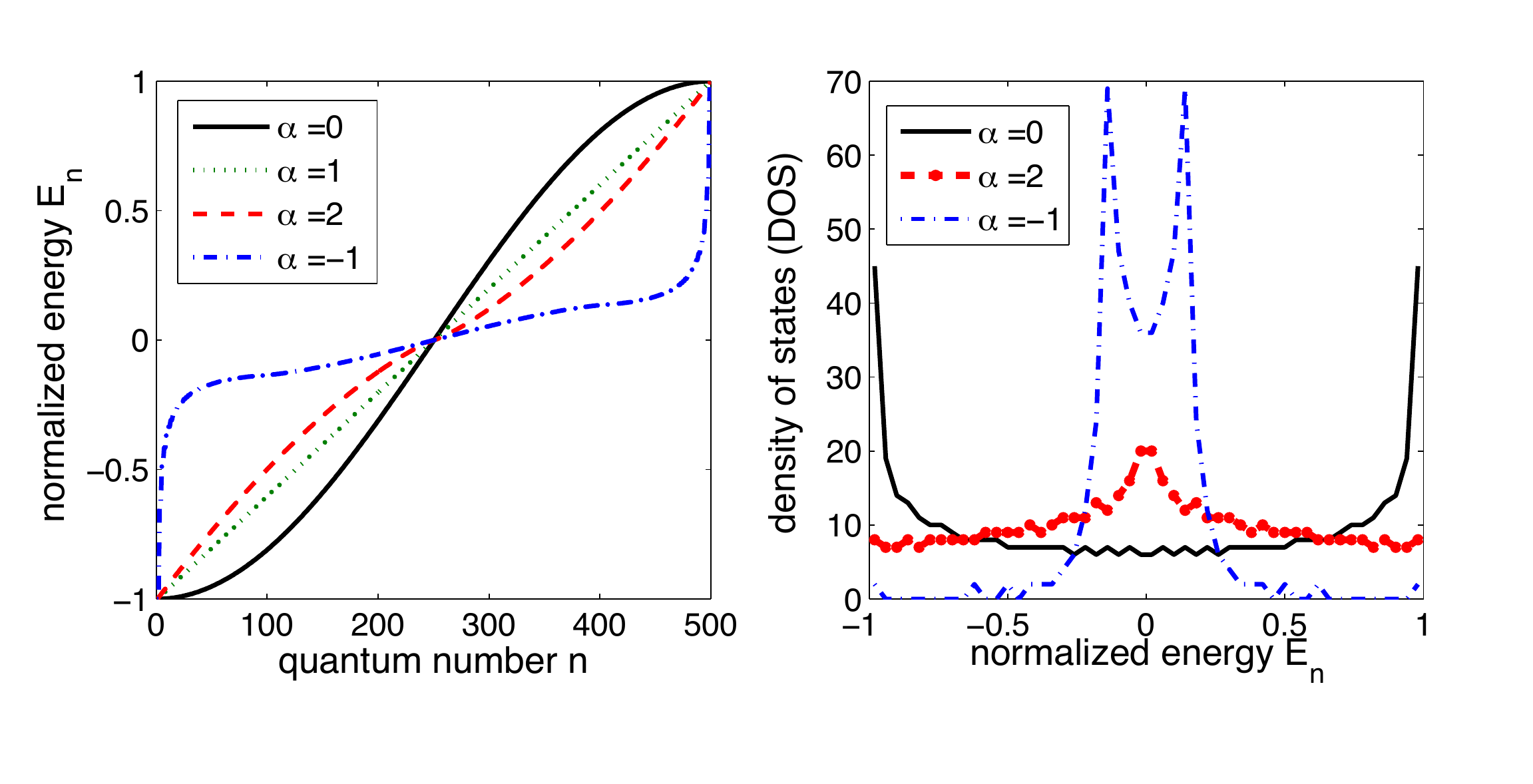}
\vspace{-1cm}
\caption{(color online) (a) Left panel shows energy spectra for the robust $\mP\mT$-symmetric chain, Eq.(\ref{eq:hpt}), with $N=500$ sites and a position-dependent hopping amplitude $t_k=t_0 k^\alpha$ with $\alpha=\{0,1,2,-1\}$. The energy is normalized by its maximum value. When $\alpha=0$ (black solid line), we recover the well-known tight-binding chain dispersion $E_n/(2t_0)=-\cos(k_n)$. When $\alpha=1$ (green dotted line), we get a linear spectrum. When $\alpha>1$ (red dashed line), the energy spectrum develops an inflection point at zero energy. In contrast, when $\alpha=-1$ (blue dot-dashed line), the energy spectrum is linear at the origin, has a steep slope near the band extrema, and develops two symmetrical inflection points. (b) Right panel shows corresponding (un-normalized) densities of states $\rho_\alpha(E)$. When $\alpha=2>1$ (red dashed line), $\rho_{\alpha=2}(E)$  develops a maximum at zero energy and it monotonically decreases to a finite value at the band edges. When $\alpha=-1< 0$ (blue dot-dashed line), the density of states shows a two-peak structure. When $\alpha=0$ (black solid line), we recover the well-known result $\rho_0(E)$ that diverges at the band edges. These results show that the energy spectrum and density of states are widely tunable through the exponent $\alpha$.}
\label{fig:band}
\end{center}
\end{figure}

\noindent{\it Energy Spectrum and Density of States:} We start with numerical results for an $N=500$ site chain with hopping amplitude given by $t_k= t_0 k^\alpha$ where $t_0$ sets the hopping-energy scale. When $\alpha=0$, we have a uniform tight-binding chain, the energy spectrum is given by $E_n= -2t_0\cos(k_n)$ where $k_n=n\pi/(N+1)$ for an open chain, and the density of states $\rho_0(x)=\theta(1-|x|)/2\pi t_0 \sqrt{1-x^2}$ diverges near the band edges $x=\pm 1$, where $x=E/(2t_0)$ and $\theta(x)$ is the Heaviside function. The left panel in Fig.~\ref{fig:band} shows the  cosine-spectrum for $\alpha=0$ (black solid line), a linear spectrum that is obtained when $\alpha=1$ (green dotted line), and nonlinear spectra obtained when $\alpha=2$ (red dashed line) and $\alpha=-1$ (blue dot-dashed line). As is expected for a tight-binding model, the energy spectra are symmetric around zero~\cite{staggered}. {\it These results show that the energy spectrum of the $\mP\mT$-symmetric chain can be widely tuned.}  We note that when $\alpha<0$, the eigenstates near the top and the bottom of the energy band are localized at the two ends of the chain. 

The right panel in Fig.~\ref{fig:band} shows the corresponding (un-normalized) densities of states $\rho_{\alpha}(E)$. It is clear that the density of states changes dramatically from $\rho_0(E)$ (black solid line) when $\alpha\neq 0$. When $\alpha=1$, due to the linear spectrum, the density of states is constant. It develops a single peak at $E=0$ and tapers off to a finite value at the band edges when $\alpha>1$. In contrast, when $\alpha<0$, it develops two symmetrical peaks and vanishes at the band edges for $N\rightarrow\infty$. We emphasize that when $\alpha\neq 0$, the system is not  translationally invariant and therefore the quantum number $n$ is not associated with the momentum.  

We now focus on $\alpha=1$ or equivalently $t_k=t_0 k$ for $k=\{1,\ldots, N-1\}$. The band-edges in this case are given by $\pm E_0=\pm (N-1)t_0$ and the uniform level-spacing is $\Delta E= E_{n+1}-E_n=2t_0$. It follows from Eq.(\ref{eq:hpt}) that the recurrence relation satisfied by the coefficients of an eigenfunction $|\psi_\gamma\rangle=\sum_{k=1}^N f_k^{\gamma} |k\rangle$ of $H_{PT}$ is
\begin{equation}
\label{eq:rr}
-t_0\left[ k f^{\gamma}_{k+1}+(N+1-k) f^{\gamma}_{k-1}\right] =E_\gamma f^{\gamma}_k.
\end{equation}
It is easy to check that $f^G_k=C^{N-1}_{k-1}=(N-1)!/(k-1)!(N-k)!$ satisfies Eq.(\ref{eq:rr}) with eigenvalue $E_G=-(N-1)t_0$. Thus the ground-state wavefunction is $|\psi_G\rangle=\sum_{i=1}^{N} f^G_k |k\rangle$. Note that for $k\sim N/2\gg 1$, Stirling approximation implies that the ground-state wavefunction is Gaussian near the center of the chain, $f^G_k\sim\exp[-(k-N/2)^2/2N^2]$. The first excited-state wavefunction is given by $f^{1}_k= (N+1-2k) f^{G}_k$. It has energy $-(N-3)t_0$, 
and Stirling approximation shows that in the large $N$-limit, it carries over to the wavefunction for the first excited state of a simple harmonic oscillator. It is straightfoward, but tedious, to construct the higher excited states. We emphasize that for every eigenstate with energy $-E<0$, the eigenstate with energy $+E>0$ is given by its staggered version: $f_k\rightarrow (-1)^k f_k$~\cite{staggered}. 

When $\alpha\neq\{0,1\}$ an analytical solution for the eigenvalue spectrum $H_{PT}$, or equivalently $H$, is unknown. However, the results in Fig.~\ref{fig:band} for $\alpha>0$ can be qualitatively understood with the simplest example of a non-trivial, symmetric, tridiagonal matrix $H$ with real entries $\{a,b,b,a\}$ above the diagonal. The matrix $H$ is similar to a $\mP\mT$-symmetric Hamiltonian $H_{PT}$ of a 5-site chain with hopping parameters $\{t_1,t_2,t_3,t_4\}$ with $a=-\sqrt{t_1t_4}$ and $b=-\sqrt{t_2t_3}$. The eigenvalues of such a matrix are given by $E_n=\{\pm\sqrt{a^2+ 2b^2},\pm a, 0\}$. For a position dependent hopping $t_k=t_0 k^\alpha$, when $\alpha=0$ the slope of the energy spectrum at the band-edge is smaller than that at the origin, when $\alpha=1$ we get the linear spectrum, and for $\alpha>1$, the slope of the spectrum at the band-edge is larger than that at the origin. 

\noindent{\it Localized and Extended Wavefunctions:} The $\mP\mT$-symmetric chain with a position-dependent hopping is not translationally invariant, and when $\alpha\neq\{0,1\}$ its eigenfunctions are not analytically known. To study the evolution of the spatial extent of a wavefunction $|\psi\rangle=\sum_{i=1}^{N}f_i |i\rangle$ with increasing system size $N$, we calculate the inverse participation ratio (IPR)~\cite{wegner}
\begin{equation}
{\rm IPR}_\psi (N)=\frac{\sum_{i=1}^{N} |f_i|^4}{\left(\sum_{i=1}^{N}|f_i|^2\right)^2}.
\end{equation}
If the IPR remains finite as $N\rightarrow\infty$, the wavefunction is localized whereas for an extended state, ${\rm IPR}(N)\propto 1/N^\eta\rightarrow 0$ where $\eta>0$; for a uniform tight-binding chain, $\alpha=0$, the ${\rm IPR}(N)=3/N$ for {\it all eigenstates}. Note that the IPRs for eigenstates with energies $\pm E$ are the same. 
\begin{figure}[h!]
\begin{center}
\includegraphics[angle=0,width=17cm]{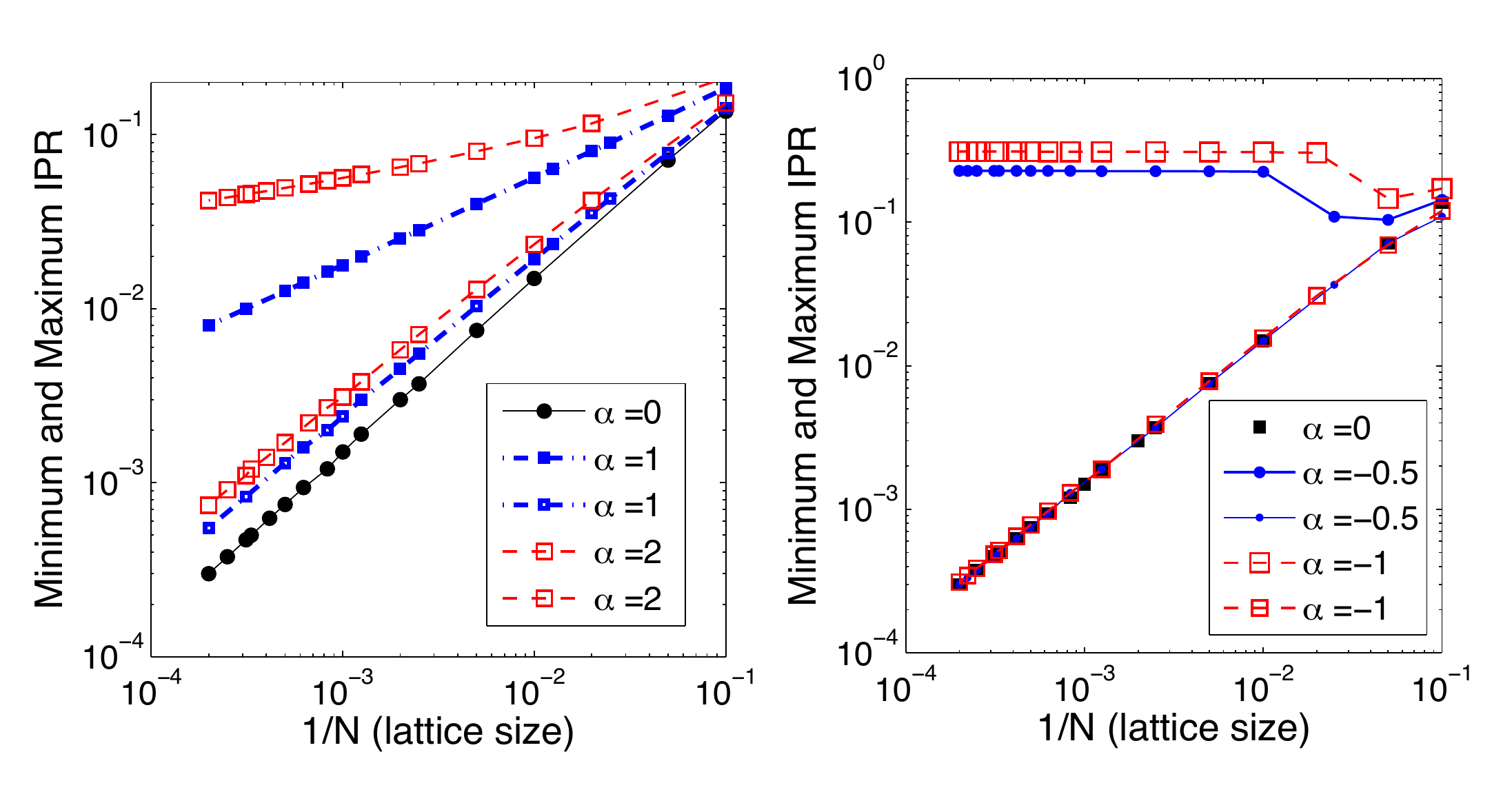}
\vspace{-1cm}
\caption{(color online) (a) The left-panel shows the minimum and maximum values of inverse-participation-ratio (IPR) for an $N$-site chain with Hamiltonian $H$, Eq.(\ref{eq:hh}), as a function of position-dependent hopping $t_k=t_0 k^\alpha$.  When $\alpha=0$ (black solid circles), we obtain the analytical result IPR$(N)\propto 1/N$. The $\alpha=1$  (blue dot-dashed line) and $\alpha=2$ (red dashed line) results show that the IPRs decrease monotonically with increasing chain size. These results strongly suggest that all eigenstates are extended when $\alpha\geq 0$. (b) The right-panel shows the IPR results for $\alpha\leq 0$. The $\alpha=-1/2$ (blue solid circles) and $\alpha=-1$ (red open squares) results show that the minimum IPR is essentially independent of $\alpha$. The maximum IPR saturates to a nonzero value and indicates the presence of localized eigenstates with energies $\pm E$. These results show that when $\alpha<0$, the system has both extended and localized states.}
\label{fig:ipr}
\end{center}
\end{figure}
The left-panel in Fig.~\ref{fig:ipr} shows the evolution of the maximum and minimum values of IPR for a  chain with $N=10$-$5000$ as a function of $\alpha\geq 0$. Note that since the chain size and the IPRs span decades, we use the logarithmic scale in Fig.~\ref{fig:ipr}. When $\alpha=0$ (black solid circles) we obtain the analytical result, ${\rm IPR} (N)=3/N$. When $\alpha>0$, we see that both the minimum and maximum IPRs decay with increasing chain size, max IPR $\propto N^{-\eta_\alpha}$ and min IPR $\propto N^{-\gamma_\alpha}$ where $0<\eta_\alpha< \gamma_\alpha<1$, and both exponents $\eta_\alpha$ and $\gamma_\alpha$ are monotonically decreasing functions of $\alpha$. These results strongly suggest that all eigenstates of the Hamiltonian $H$ with position-dependent hopping $t'_k = t_0\left[k(N-k)\right]^{\alpha/2}$ are extended when $\alpha\geq 0$. The right panel shows corresponding results for $\alpha\leq 0$. The minimum IPR is essentially independent of $\alpha$. On the other hand, in a sharp contrast with the previous results, we see that the maximum IPR quickly saturates to a nonzero value and indicates a localized state. Thus, when $\alpha<0$, {\it the system has both extended and localized eigenfunctions}. We note that these exponentially localized states, at the two ends of the chain, are essentially degenerate in energy; so are their staggered counterparts~\cite{staggered}. Thus, there are at least four eigenstates that have the same nonzero IPR. The qualitative difference between $\alpha>0$ and $\alpha<0$ cases can be attributed to the hopping: when $\alpha>0$, the hopping amplitude {\it increases} from $\sim t_0 N^{\alpha/2}$ at the two edges to $\sim t_0 N^{\alpha}$ at the center of the chain, whereas when $\alpha<0$, the hopping amplitude {\it decreases} from $\sim t_0/N^{|\alpha|/2}$ at the edges to $\sim t_0/N^{|\alpha|}$ at its center.

\begin{figure}
\begin{center}
\includegraphics[angle=0,width=13cm]{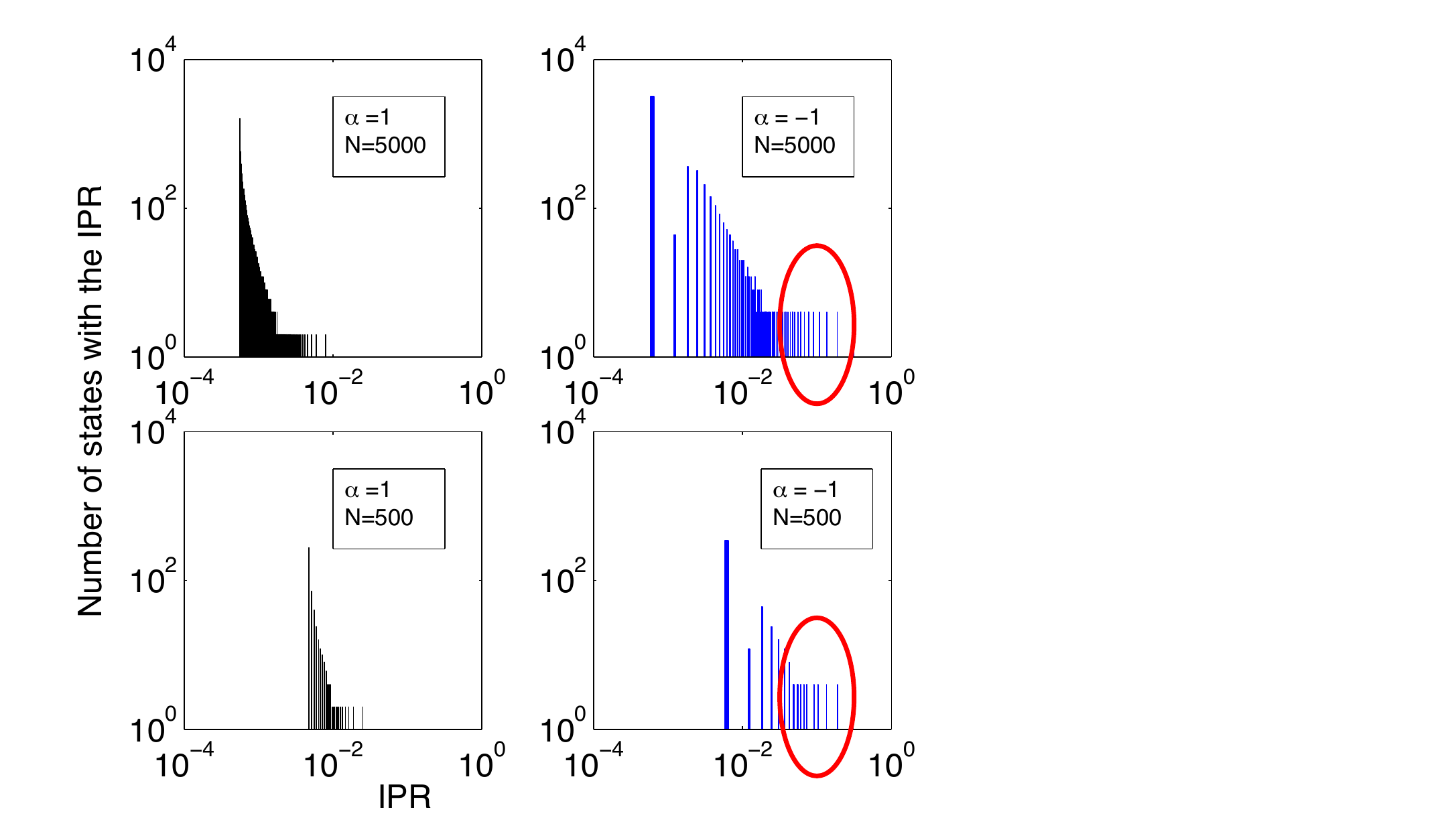}
\caption{(color online) The evolution of the inverse-participation-ratio (IPR) distribution with the chain size $N$. Note the logarithmic scale. The top left and bottom left panels show that when $\alpha=1$, as $N$ increases, the entire distribution of IPRs shifts to lower values. It suggests that all eigenstates of the Hamiltonian $H$, Eq.(\ref{eq:hh}), are extended in the absence of disorder. The top right and bottom right panels show that when $\alpha=-1$, as $N$ is increased, although most of the IPRs shift to lower values, they saturate to a nonzero value for the states shown in the red oval. These eigenstates are localized at the two ends of the chain, and each finite value of the IPR is four-fold degenerate. Thus, when $\alpha<0$, the system has both extended and localized eigenstates in the absence of disorder.}
\label{fig:iprdist}
\end{center}
\end{figure}

A better insight into the number of localized states is provided by the dependence of the IPR distribution on the size $N$ of the chain.  Figure~\ref{fig:iprdist} shows the histogram of IPRs for $\alpha=1$ (left column) and $\alpha=-1$ (right column). When $\alpha=1$, as $N$ is increased tenfold from $N=500$ (bottom-left panel) to $N=5000$ (top-left panel), the entire IPR distribution shifts to smaller values. In contrast, when $\alpha=-1$, even as $N$ is increased tenfold, the IPR values for a few (localized edge) states, indicated by the red oval, are unchanged while the IPRs for the rest shift to lower values. 

\noindent{\it Discussion:} $\mP\mT$-symmetric lattices with a uniform hopping and imaginary impurities have an extremely fragile $\mP\mT$-symmetric phase~\cite{bendix,mark}. In this paper, we have presented a tight-binding model with non-Hermitian, $\mP\mT$-symmetric hopping, Eq.(\ref{eq:hpt}). We have shown that, under very general circumstances, {\it this model is always in the $\mP\mT$-symmetric phase}, and its Hamiltonian is similar to a Hermitian Hamiltonian with position-dependent nearest-neighbor hopping, Eq.(\ref{eq:hh})~\cite{gautam}. These results are unaffected by the presence of ubiquitous, on-site, Hermitian disorder since it does not induce $\mP\mT$-symmetry breaking. 

Given the robust nature of the $\mP\mT$-symmetric phase in this chain, we have explored its energy spectrum, density of states, nature of eigenfunctions, and their dependence on the functional form of the hopping amplitude $t_k=t_0 k^\alpha$. We find that when $\alpha=1$ the energy spectrum is linear and gives rise to a constant density of states. We show that the energy spectrum is widely tunable by changing $\alpha$. We find that when $\alpha<0$ the system has both localized and extended eigenfunctions in the absence of disorder, whereas when $\alpha>0$, all eigenfunctions are extended. The effect of a Hermitian on-site disorder, then, is identical to that in a regular tight-binding model~\cite{anderson,wegner}.   Thus, the physics of the robust $\mP\mT$-symmetric chain with non-Hermitian hopping is extremely rich. 


Y.J. acknowledges useful discussions with Donald Priour and Ricardo Decca. This work was supported in part by the U.S. Department of Energy. 


\end{document}